\documentclass[useAMS,usenatbib,twocolumn]{mn2e}
\usepackage{times} 
\usepackage{amsmath}
\usepackage{rotating}
\usepackage{amssymb}

\def\aap{A\&A}
\def\apj{ApJ}

\def\mnras{MNRAS}

\def\apjs{ApJS}
\def\prd{Phys. Rev. D}
\usepackage{color}

\title[Nonspherical similarity solutions for dark halo formation]
{Nonspherical similarity solutions for dark halo formation}
\author[Vogelsberger et al.] 
{
Mark Vogelsberger$^1$\thanks{mvogelsberger@cfa.harvard.edu}, 
Roya Mohayaee$^2$, Simon D.M. White$^3$ 
\\
(1)
Harvard-Smithsonian Centre for Astrophysics, 60 Garden Street, MA 02138, USA\\
(2) 
Institut d'Astrophysique de Paris (IAP), CNRS, UPMC, 
98 bis boulevard Arago, France\\
(3) 
Max-Planck Institut fuer Astrophysik,
Karl-Schwarzschild Strasse 1, 
D-85748 Garching, 
Germany 
}

\begin{document}
\date{Accepted ???. Received ???; in original form ???}

\pagerange{\pageref{firstpage}--\pageref{lastpage}} \pubyear{2010}

\maketitle

\label{firstpage}

\begin{abstract}
We carry out fully 3-dimensional simulations of evolution from self-similar,
spherically symmetric {\it linear} perturbations of a Cold Dark Matter
dominated Einstein-de Sitter universe. As a result of the radial orbit
instability, the haloes which grow from such initial conditions are triaxial
with major-to-minor axis ratios of order 3:1. They nevertheless grow
approximately self-similarly in time.  In all cases they have power-law
density profiles and near-constant velocity anisotropy in their inner
regions. Both the power-law index and the value of the velocity anisotropy
depend on the similarity index of the initial conditions, the former as
expected from simple scaling arguments. Halo structure is thus not
``universal'' but remembers the initial conditions. On larger scales the
density and anisotropy profiles show two characteristic scales, corresponding
to particles at first pericentre and at first apocentre after infall. They are
well approximated by the NFW model only for one value of the similarity index.
In contrast, at all radii within the outer caustic the pseudo phase-space
density can be fit by a single power law with an index which depends only very
weakly on the similarity index of the initial conditions. This behaviour is
very similar to that found for haloes formed from $\Lambda$CDM initial
conditions and so can be considered approximately universal.
\end{abstract}


\begin{figure*}
\center
{
\includegraphics[width=0.49\textwidth]{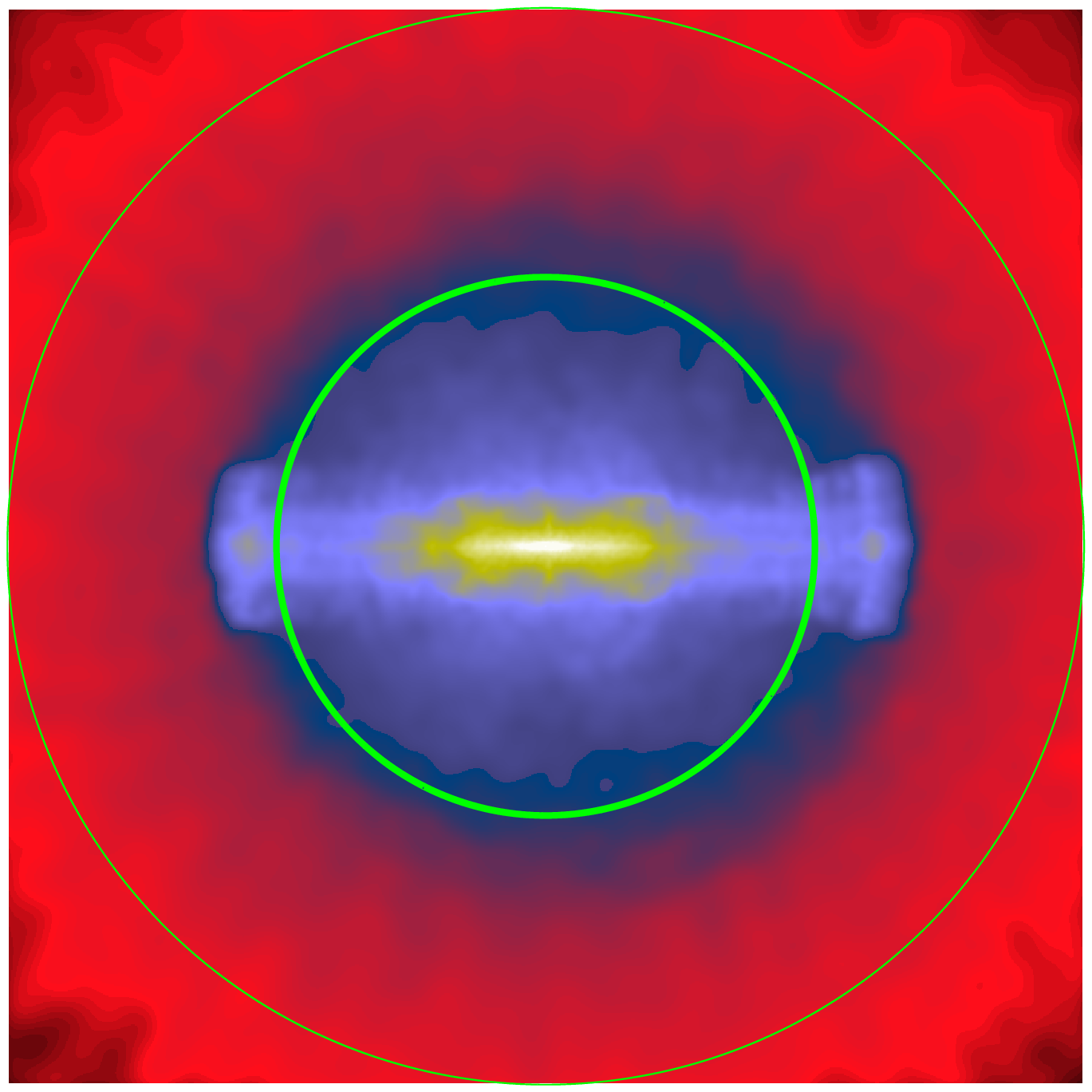}
\includegraphics[width=0.49\textwidth]{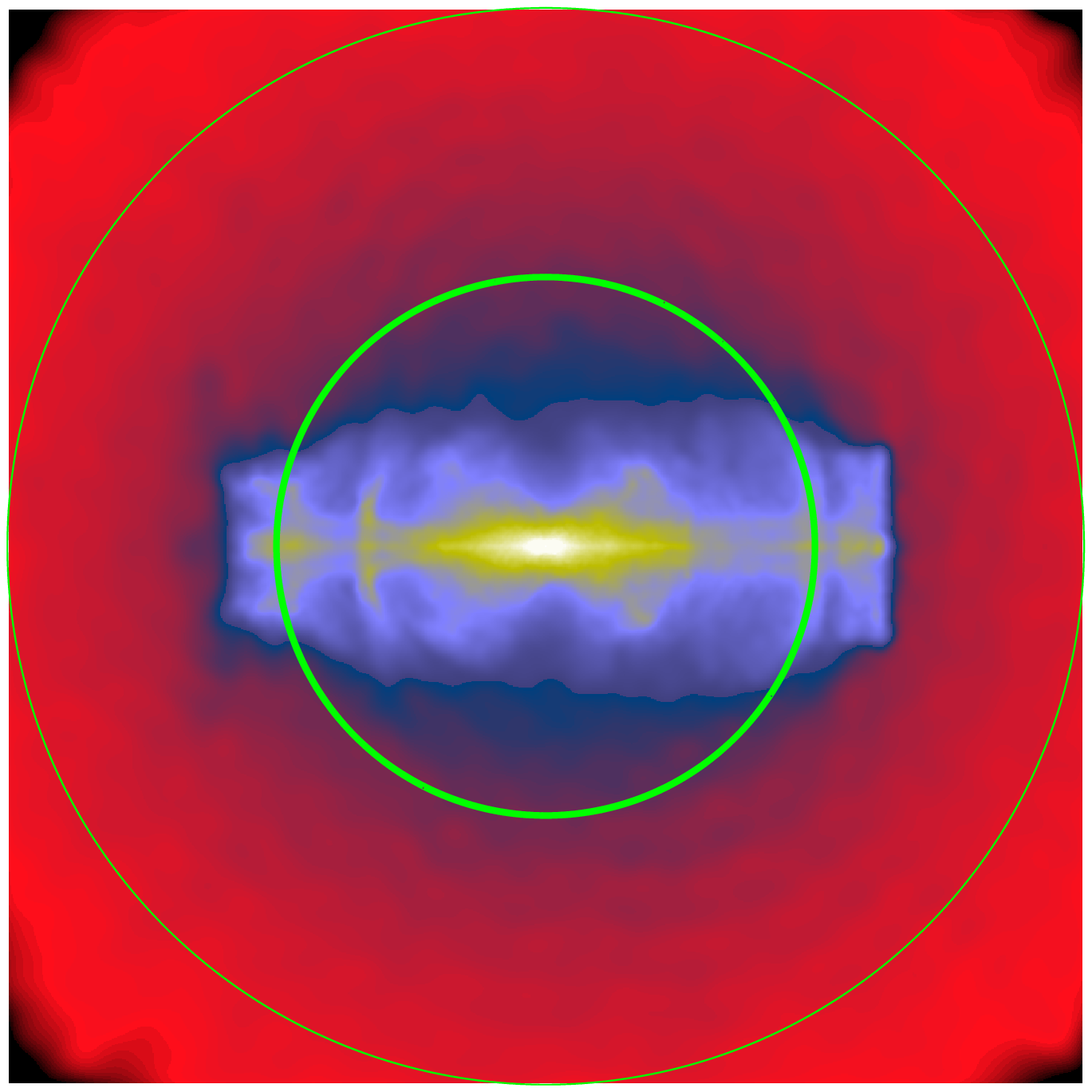}
}
\caption{ Projected density maps of the haloes formed for
  $\epsilon=0.4$ (left) and $\epsilon=0.8$ (right).  In each case, the
  long axis of the bar is horizontal, its short axis is vertical, and
  the region plotted is a slice of thickness $2r_{\rm outer\,\, caustic}$ 
  and side  $4r_{\rm outer\,\, caustic}$ where $r_{\rm outer\,\, caustic}$ 
  is the radius of the outer caustic (first apocentre after the turnaround) 
  in the spherical similarity solution.  The thick and thin green circles have 
  radii of $r_{\rm outer\,\, caustic}$ and $2r_{\rm outer\,\, caustic}$
  respectively. In units of the outer-caustic radius, the bar length
  is approximately ``universal'', although time-dependent features are
  still visible in these plots, for example, in the lack of left-right
  symmetry.}
\label{fig:density_map} 
\end{figure*}

\section{Introduction}
\label{section:introduction}

Galaxies are believed to form from gas condensing at the centres of massive
dark matter haloes as these grow by collapse and aggregation from weak density
fluctuations emerging from the early Universe \citep{1978MNRAS.183..341W}.

The earliest theoretical insights into the formation and evolution of dark
matter haloes were provided by the spherical infall model of
\cite{1972ApJ...176....1G} and \cite{1975ApJ...201..296G}.  In this model, an
isolated overdensity in an otherwise unperturbed Einstein-de Sitter universe
first expands with the Hubble flow, then turns around and
collapses. Surrounding material continues to fall onto the object until its
mass greatly exceeds that of the originally perturbed region. Asymptotically,
a power-law density profile is established with $\rho\propto r^{-2.25}$. The
late-time structure of this model is a similarity solution whose structure was
worked out in detail by \cite{1985ApJS...58...39B}. More general similarity
solutions where the initial mass perturbation scales with enclosed mass as
$\delta M/M\propto M^{-\epsilon}$ were presented by \cite{1984ApJ...281....1F}
who showed that these produce haloes with $\rho\propto r^{-\gamma}$, where
$\gamma = 2$ for $0 < \epsilon \leq 2/3$ and $\gamma =
9\epsilon/(1+3\epsilon)$ for $\epsilon \geq 2/3$. The change in behaviour at
$\epsilon=2/3$ reflects the fact that all orbits are purely radial in these
models. No superposition of radial orbits can self-consistently produce a
power-law shallower than $r^{-2}$ at small $r$. Further generalisations to
include randomly oriented rosette orbits with a scale-free eccentricity
distribution resulted in spherically symmetric similarity solutions for which
$\gamma = 9\epsilon/(1+3\epsilon)$ for all $\epsilon > 0$
\citep{1992ApJ...394....1W,1995PhRvL..75.2911S,
  1997PhRvD..56.1863S,2001MNRAS.325.1397N,2006ApJ...653...43M}. In such models
the inner density structure of the final haloes reflects the scaling
properties of the initial conditions.

In contrast, N-body simulations have shown for more than a decade that dark
matter haloes formed from fully 3-dimensional, cosmologically realistic
initial conditions do not have pure power-law density profiles. Rather, the
logarithmic slope of simulated profiles changes slowly but continuously with
radius. In addition, the shape of these profiles is almost independent of halo
mass, of cosmological parameters, and of the power spectrum of initial
fluctuations \citep{1997ApJ...490..493N}.  The most popular representation of
this ``universal'' shape is the NFW model which behaves as $1/r$ in the inner
regions and as $1/r^3$ in the outskirts \citep{1997ApJ...490..493N}.  Even
haloes formed by monolithic collapse in the first nonlinear phases of hot and
warm dark matter models (HDM and WDM) are well represented by NFW fits,
showing that hierarchical growth is not required to produce this universality
(see {\it e.g.}~\cite{1999ApJ...517...64H, 2009MNRAS.396..709W}). Recent N-body
simulations with significantly increased numerical resolution have found small
but significant deviations from NFW shape which depend systematically on halo
mass \citep{2008MNRAS.388....2H,2010MNRAS.402...21N}. Thus halo density profiles are not
truly universal.  Nevertheless, these variations are much smaller than those
predicted by the similarity solutions, and the profiles can be considered
universal to a good approximation.

The differences in behaviour between similarity solutions and numerical
simulations reflect the fact that the former enforce spherical symmetry and a
potential which varies smoothly with time according to the similarity scaling,
while the latter involve strongly time-dependent and fully three-dimensional
potential fluctuations which exchange energy and angular momentum between
different parts of the system through the classical ``violent relaxation''
mechanism \citep{1967MNRAS.136..101L}. Even though extensions of the original
similarity solutions allow non-radial orbits with a variety of eccentricity
distributions \citep{1992ApJ...394....1W,1995PhRvL..75.2911S,
  1997PhRvD..56.1863S,2001MNRAS.325.1397N,2006ApJ...653...43M}, they retain
both spherical symmetry and strict similarity scaling, and so exclude violent
relaxation and any possibility for it to drive convergence towards a universal
(i.e. $\epsilon$-independent) nonlinear structure.  Large and time-dependent
potential fluctuations are required to produce such convergence. These occur
naturally in hierarchical assembly models but also in other situations, for
example, during the monolithic, quasi-ellipsoidal collapse of the first
generation of haloes in HDM or WDM cosmogonies \citep{1970A&A.....5...84Z,
2001MNRAS.323....1S,2009MNRAS.396..709W}.

\begin{figure*}
\center
{
\includegraphics[width=0.49\textwidth]{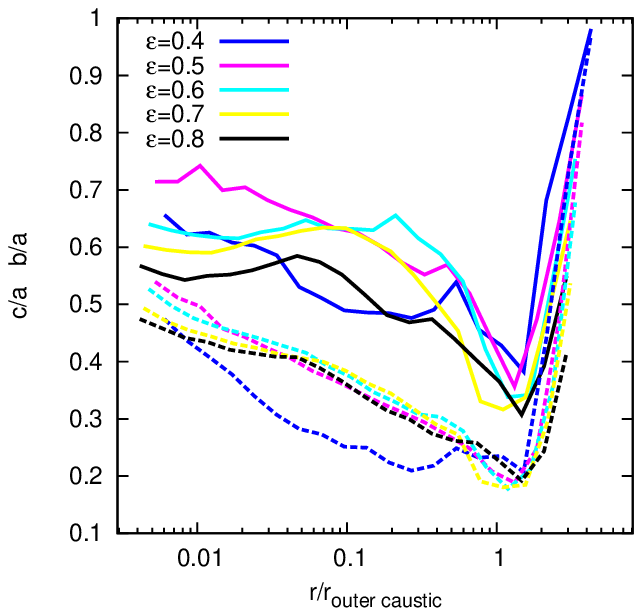}
\includegraphics[width=0.49\textwidth]{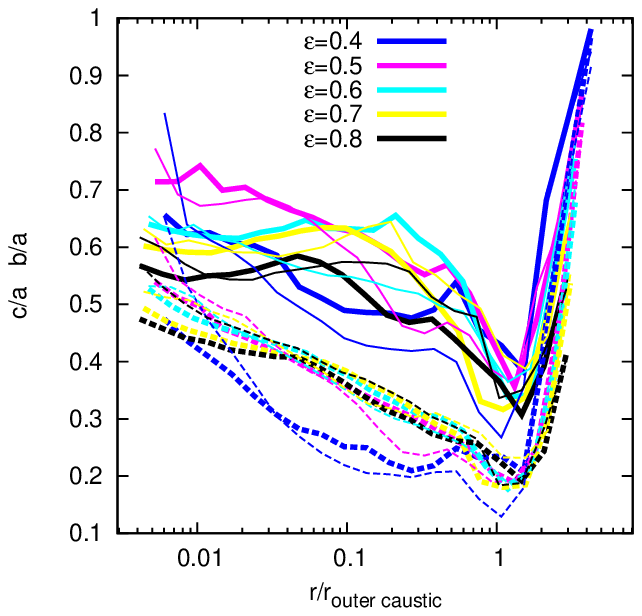}
}
\caption
{ Left panel: Halo axis ratios as a function of radius for a variety of
  similarity scaling parameters. Different colours correspond to different
  values of the scaling parameter $\epsilon$.  For each $r$, axis ratios are
  estimated from the principal values of an inertia tensor calculated for all
  particles within an ellipsoid of mean radius $r$. Dashed lines denote
  minor-to-major axis ratios while solid lines denote intermediate-to-major
  ratios. For each simulation, radii are normalised by that of the outer
  caustic in the corresponding similarity solution.  The left panel shows
  results for the final time $a=1$. In the right panel these are replotted as
  thick lines and compared with results for $a=0.5$ (the thin lines). }
\label{fig:axisratio} 
\end{figure*}
 
It has long been known that spherical equilibria dominated by radial orbits
are violently unstable and evolve on a few dynamical times into strongly
ellipsoidal bars with significant nonradial motions
\citep{1973dgsc.conf..139A,1973A&A....24..229H,1981SvAL....7...79P,
  1981AZh....58..933P,1985IAUS..113..297B,1985MNRAS.217..787M,1997ApJ...490..136M}.
The original similarity solutions of \cite{1984ApJ...281....1F} and
\cite{1985ApJS...58...39B} are thus not viable models for the formation of real
systems -- the slightest non-spherical perturbation of their quasi-equilibrium
inner regions causes rapid evolution to an entirely different nonlinear
structure. A number of authors have noted situations where the radial orbit
instability gives rise to haloes with NFW-like density profiles
\citep{1999MNRAS.302..321H,1999ApJ...517...64H,
  2005ApJ...634..775B,2006ApJ...653...43M,2008ApJ...685..739B}. There are thus
(at least) two possibilities for the long-term evolution of structure from
(almost) spherically symmetric, self-similar linear initial conditions.
Either it may approach a non-spherical self-similar solution, which would then
have a potential which changes smoothly in time and no violent relaxation, or
the strongly time-dependent behaviour may continue indefinitely, allowing
violent relaxation to rearrange material in the inner regions. In the former
case an NFW-like universal profile is only consistent with similarity scaling
for $\epsilon = 1/6$, whereas in the latter case an NFW-like profile could,
in principle, be maintained at all times.

In the present paper we investigate these issues by simulating evolution from
spherically symmetric, self-similar, linear initial conditions for a variety
of values of $\epsilon$.  As we showed in \cite{2009MNRAS.tmp.1490V} particle
noise leads to the rapid onset of the radial orbit instability in such
simulations, so that their later nonlinear evolution is strongly
non-spherical. Here we show that while some chaotic time-dependence remains at
late times, evolution is nevertheless approximately self-similar. The inner
structure of the haloes remembers the initial conditions from which they
formed, depending on $\epsilon$ in the same way as in the spherical similarity
solutions.  In Section \ref{section:bar} we present our simulations and study
the formation of ellipsoidal ``bars''.  In Section \ref{section:density} we study
how these bars affect the velocity anisotropy and density profiles in the
inner halo.  The transition between the infall and quasi-equilibrium regions,
and the inner and outer scales which define it are discussed in Section
\ref{section:2scales}.  Finally in Section \ref{section:psd} we demonstrate
that, in contrast to the velocity anisotropy and density profiles, the profile
of pseudo-phase-space density is almost universal in these models.

\section{Radial-orbit instability: bar formation} 
\label{section:bar}

The similarity solutions assume an Einstein-de Sitter
universe in which the linear mass perturbation $\delta M_i$ within a
sphere containing unperturbed mass $M_i$, initially satisfies 
\begin{equation}
\frac{\delta M_i}{M_i} = 1.0624 \left(\frac{M_i}{M_0}\right)^{-\epsilon},
\label{eq:initial-profile}
\end{equation}
where $\epsilon$ is a scaling index and $M_0$ is a reference mass taken to be
the mass within the turnaround radius ($r_{\rm ta}$) at the initial time.  The
parameter $\epsilon$ is restricted to positive values in order to ensure that
more distant mass shells turn around and fall back later than inner ones. The
mass within the shell that is just turning around and the physical radius of
this shell at turnaround then scale with time as $M_{\rm ta}\propto
t^{2/3\epsilon}$ and $r_{\rm ta}\propto t^{2/3 + 2/9\epsilon}$ so that the
mean mass density contained within the turnaround radius is always 5.5 times
the cosmic mean (the critical density) and so satisfies $\rho_{\rm ta} \propto
t^{-2}$.

\begin{figure*}
\center{
\includegraphics[width=0.49\textwidth]{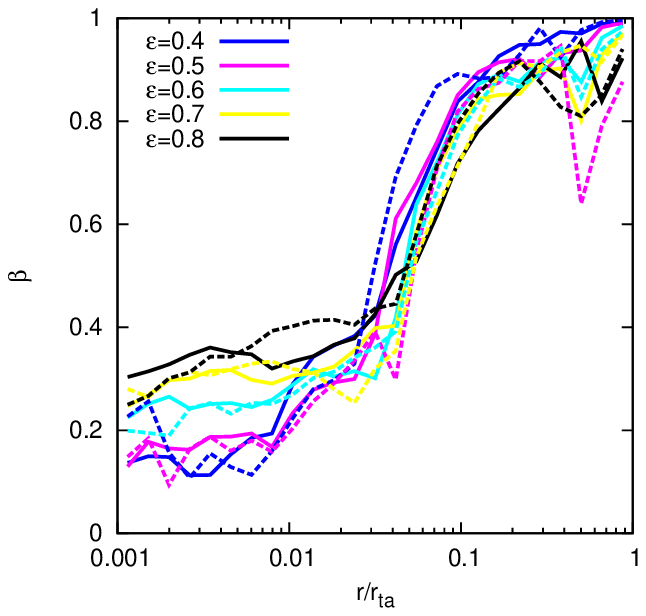}
\includegraphics[width=0.475\textwidth]{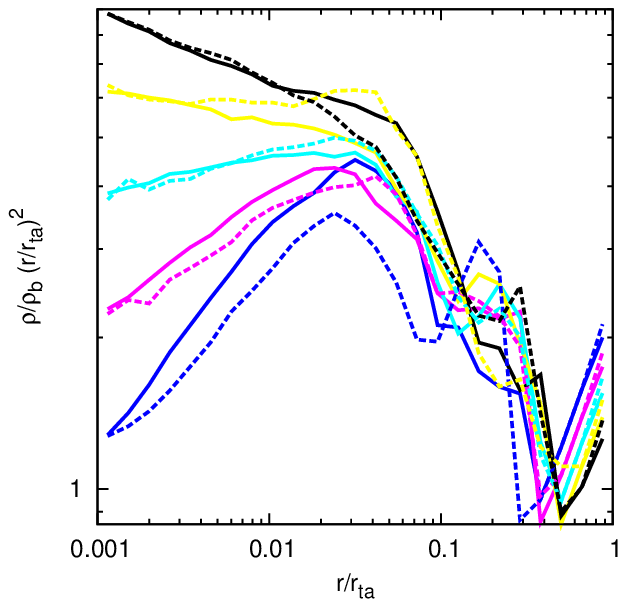}
}
\caption{Left panel: Velocity anisotropy $\beta(r)=1-\sigma_t^2/\sigma_r^2$ as
  a function of radius $r$, normalised by the turnaround radius $r_{\rm
    ta}(t)$.  Here $\sigma_r(r)$ and $\sigma_t(r)$ are the 1-dimensional
  radial and transverse velocity dispersions averaged over spherical
  shells. Different colours denote simulations with different values of
  $\epsilon$ as indicated. Solid lines are for $a=1$ and dashed lines for
  $a=0.5$. Beyond the outer caustic, all particles are falling in for the
  first time, but they nevertheless have nonzero radial and tangential
  velocity dispersions because the quadrupole moment of the ellipsoidal inner
  regions induces variations in infall velocity and infall direction at each
  $r$. Both dispersions jump dramatically as the outer caustic is crossed and
  the anisotropy is slowly varying and relatively small in the inner
  quasi-equilibrium regions.  The softening radius is too small to affect the
  velocity dispersions over the radial range plotted.  Right panel: Mass
  density averaged in spherical shells as a function of radius, again
  normalised by $r_{\rm ta}$. Line colours and types correspond to those in
  the left panel.  The density is given in units of the current critical
  density and is multiplied by $(r/r_{\rm ta})^2$ to suppress the dominant
  dependence and to allow easier comparison of the curves. In both panels the
  structure is very similar at the two times considered. The dependence on
  $\epsilon$ is very clear in the density profiles and appears significant
  also in the anisotropy profiles.}
\label{fig:half_time_profiles} 
\end{figure*}

Since these initial conditions are spherically symmetric, the system
remains spherically symmetric at later times (if instabilities are
ignored/suppressed) and the particles on each spherical shell all execute
identical though differently oriented radial orbits. The scale-free nature
of the initial conditions ensures that the orbits executed by different shells
are identical when scaled to their individual turnaround radii and times.
\cite{1972ApJ...176....1G} showed that at radii much smaller than $r_{\rm ta}$ this
results in a stable halo with density profile
$\rho\propto r^{-9/4}$ when $\epsilon = 1$, the only case they
considered. \cite{1984ApJ...281....1F} generalised this to other values
of $\epsilon$, finding the inner profile to vary with the initial
scaling properties:
\begin{equation}
\rho(r) \propto \left\{
\begin{array}{lcl}
r^{-2}                       & \qquad , \qquad & \epsilon\le {2\over 3} \\ \\
r^{-9\epsilon/(1+3\epsilon)} & \qquad , \qquad & \epsilon\ge {2\over 3}.\\
\end{array}
\right.
\label{eq:profile0}
\end{equation}
The change of behaviour at $\epsilon=2/3$ is due to the fact that no
self-consistent spherical equilibrium system built from radial orbits can have
a power-law slope shallower than $-2$, because every particle is constrained
to pass through the centre of the system once per orbit. The full structure of
the similarity solution for the case $\epsilon=1$ was worked out by
\cite{1985ApJS...58...39B}, who showed that the power-law inner structure breaks
in the transition to the infall regime, and that the confinement of non-zero
phase-space density to a 3-dimensional sheet results in a series of sharp
spherical caustics which are superposed on the inner power-law density profile
and are located at any given time at the positions of particles currently
passing through apocentre.

In this paper we simulate evolution from initial conditions which obey the
spherical similarity solutions, exactly as in \cite{2009MNRAS.tmp.1490V}, but
for a variety of values of $\epsilon$.  We use a fully three-dimensional
N-body solver, a version of Gadget-2 \citep{2005MNRAS.364.1105S}, and we allow
particle noise to drive the radial orbit instability. This happens in the
first few expansion factors so that the later stages of the simulations all
contain fully developed ellipsoidal ``bars''. All our simulations follow evolution
over a factor of 1000 in time corresponding to expansion of the background
cosmology from $a=0.01$ and until $a=1$. The softening length is kept fixed in
comoving coordinates at $0.00025r_{\rm ta}(a=1)$. The final halo is
represented by about $256^3/2 \sim 8.4\times 10^6$ particles within $r_{\rm ta}$.

In Fig.~\ref{fig:density_map} we show images of the bar in the final state of
two of our simulations, $\epsilon=0.4$ on the left and $\epsilon=0.8$ on the
right. In each case the bars extend slightly beyond the position of the
outer caustic of the corresponding spherical similarity solution, which we
indicate with a green circle. The location of the actual outer caustic
corresponds to the transition in colour from blue to red. In all our
simulations we find bars with semi-major axes (delineated by the outer
caustics) which are about 1.2 times the radius of the spherical caustic. Thus
measured in units of the latter, our bar lengths appear universal. In both the
cases shown there are clear left-right asymmetries in the inner structure
of the bars (the yellow regions). Such asymmetry is a clear indicator of
time-dependent behaviour -- any exact non-spherical similarity solution
is expected to be left-right symmetric. 

We show the axis-ratios of equidensity surfaces of our final bars as a
function of semi-major axis in the left panel of Fig.~\ref{fig:axisratio}.
These ratios were obtained by calculating the moment of inertia of all
particles within an ellipsoidal surface, deriving the eigenvalues and
principal axes, and then iterating until the shape and orientation of the
bounding ellipsoid are consistent with the orientation and relative axis
lengths inferred from the moment of inertia. These ratios are thus cumulative,
referring to all material interior to the quoted radius rather to an
ellipsoidal shell at this radius. The axis ratios take their most extreme
values just inside the outer caustic, with all simulations giving
minor-to-major values close to 0.2 and intermediate-to-major values close to
0.35. They approach unity rapidly at larger radii, and also rise steadily
though more slowly towards smaller radii. There is no clear trend as a
function of $\epsilon$, suggesting that the exact values may be stochastically
determined and perhaps also time-dependent.  We test this in the right panel
by replotting the curves using a thick line-style and comparing them with
results at $a=0.5$, represented by thin lines. The overall pattern is very
similar at the two times. There is some indication that the deviation of
individual models from the mean is coherent between the two times, but there
are also substantial variations. Time-dependent effects appear to be
influencing the measurements significantly, and these plausibly have a
relatively long time coherence. In consequence it is not possible to decide
whether the apparent trends with $\epsilon$ are real, or just reflect
unrelated time-dependent variations. The overall similarity of all the curves
at both times suggest that the shape behaviour is approximately both
self-similar and ``universal''.

\section{Density and velocity anisotropy profiles}
\label{section:density}

The central bar torques infalling particles, inducing non-radial 
motions which transform the purely radial orbits of the original
similarity solution into fully three-dimensional orbits. This alters both the
density and the velocity dispersion structure of the nonlinear regions,
as we now illustrate.  A velocity
anisotropy parameter is conventionally defined as
\begin{equation}
\beta(r)=1-\sigma_t^2/\sigma_r^2,
\end{equation}
where $\sigma_r(r)$ and $\sigma_t(r)$ are 1-dimensional radial and transverse
velocity dispersions averaged over spherical shells of radius $r$.  The left
panel of Fig.~\ref{fig:half_time_profiles} shows $\beta(r)$ at two different
times, $a=0.5$ (dashed lines) and $a=1$ (solid lines). The right panel shows
$(r/r_{\rm ta})^2\rho(r)/\rho_b$ at the same two times. The two quantities are
evaluated using the same set of logarithmically spaced spherical
shells. Neither profile changes systematically with time when plotted against
$r/r_{\rm ta}$, consistent with the expectations of self-similar evolution.

The spherically-averaged density profiles in the right panel of
Fig.~\ref{fig:half_time_profiles} show three distinct regimes. At radii larger
than the outer caustic, only one (infalling) matter stream is present.
Here, there is essentially perfect agreement between $a=0.5$ and $a=1$ in all
cases, and the effective power-law index of the profile is significantly
greater than $-2$ and similar for all $\epsilon$. Only the position of the
outer caustic (in units of $r_{\rm ta}$) varies significantly with $\epsilon$.
Between this outer caustic and the typical radii of particles passing their
first pericentre after turnaround\footnote{We investigate these scales in more
  detail below.}, the density profile has a mean effective index which is
similar for all $\epsilon$ and substantially less than $-2$.  The behaviour in
this regime is quite irregular, however, and there are substantial differences
between $a=0.5$ and $a=1$ in several cases.  Examination of phase-space plots
(see, for example, Fig.~8 of \cite{2009MNRAS.tmp.1490V}) suggests that this
time-dependence is driven by large-scale irregularities which grow around
the first apocentre after turnaround but are washed out by phase-mixing
at later times. Inside the radius of first pericentre, the profiles become
more regular and agree well between the two times shown.  To a good
approximation, they are power laws that are well fitted
(as $r\rightarrow 0$) by
\begin{equation}
\rho (r) \propto r^{-9\epsilon/(1+3\epsilon)}, \quad \epsilon > 0 \\ \\.
\label{eq:profile}
\end{equation}
This is the value expected for a perfect similarity solution in the presence
of nonradial motions \citep{1992ApJ...394....1W,1995PhRvL..75.2911S,
  1997PhRvD..56.1863S,2001MNRAS.325.1397N,2006ApJ...653...43M}, demonstrating
that the time-dependent behaviour seen at intermediate radii is not sufficient
to destroy the similarity scaling and enforce ``universal'' structure. Note
that both the steep slope over the decade immediately inside the outer caustic
radius (which is roughly at the conventional virial radius) and the break to a
shallower slope at smaller radii are qualitatively similar to the ``universal''
behaviour encapsulated by the NFW profile. However, the inner slope matches
the NFW value only for $\epsilon = 1/6$, smaller than any of the values tested
here.

The left panel of Fig.~\ref{fig:half_time_profiles} shows that velocity
anisotropy profiles are similar for all values of $\epsilon$, exhibiting
distinct behaviour in the same three regimes seen in the density profiles.  At
large radii, particles are falling in for the first time on very nearly radial
orbits, gradually gaining angular momentum as they feel the quadrupole moment
of the central bar. As they cross the outer caustic they are mixed with
particles which have already passed pericentre, and $\beta(r)$ drops to much
smaller values. Inside the radius of first pericentre,  $\beta(r)$ is almost 
constant in the quasi-equilibrium region, declining slightly
towards the centre.  In this region there appears to be a weak but significant
dependence of anisotropy on similarity index, with larger values of $\epsilon$
giving rise to slightly more radially biased velocity dispersions. This
behaviour is a consequence of the dependence of the inner density profile on
$\epsilon$.  For larger $\epsilon$ the inner mass distribution is more
strongly centrally concentrated, so that its quadrupole moment become less
significant relative to the monopole and less angular momentum is induced in
the orbits of infalling particles.

\begin{figure}
\center{
\includegraphics[width=0.48\textwidth]{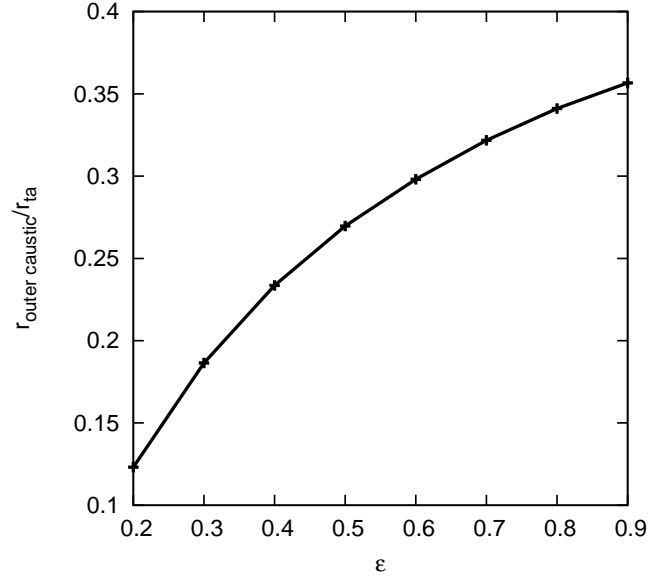}
}
\caption{ The radius of first apocentre after turnaround (i.e. the outer
  caustic radius) in units of the current turnaround radius as a function of
  $\epsilon$ in the spherically symmetric similarity solutions with radial
  orbits.}
\label{fig:caustic-ta} 
\end{figure}

\begin{figure}
\center{
\includegraphics[width=0.48\textwidth]{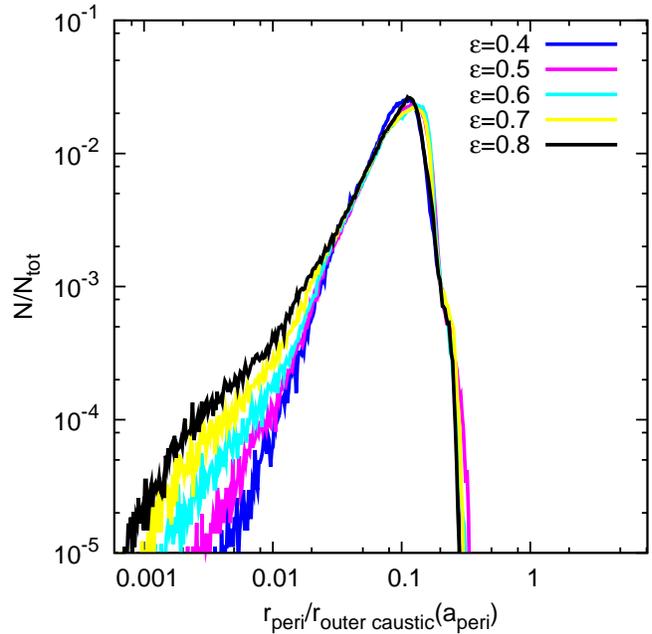}
}
\caption
{ Histograms of first pericentre distance, measured directly from our
  simulations, in units of the fiducial current outer caustic radius given in
  Fig.~\ref{fig:caustic-ta}. In these units the distribution of first
  pericentre distance is almost independent of $\epsilon$, peaking at about
  0.1.}
\label{fig:peri_distance} 
\end{figure}

\begin{figure*}
\center
{
\includegraphics[width=0.49\textwidth]{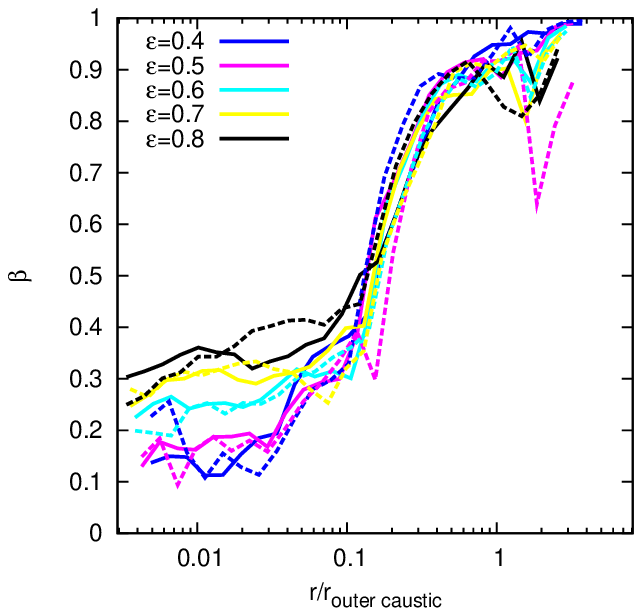}
\includegraphics[width=0.46\textwidth]{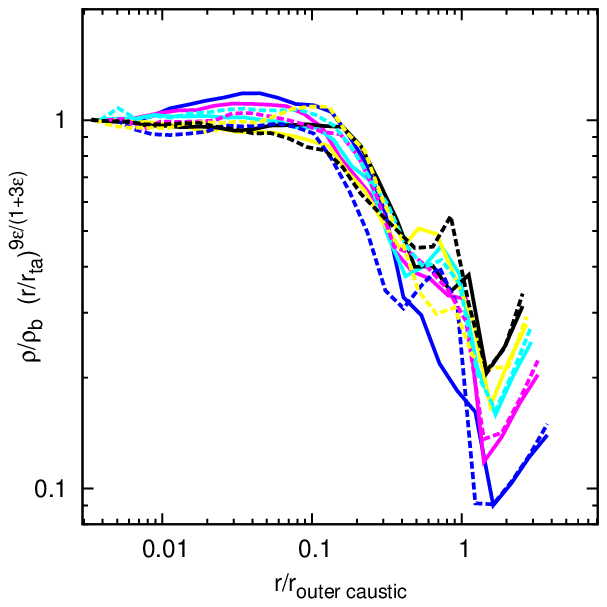}
}
\caption{Using the radius of the outer caustic as a characteristic scale
  makes the nonlinear structure of haloes appear almost 
  independent of $\epsilon$. Left and right panels show the effect of
  using this scaling on the velocity anisotropy and density profiles,
  respectively. In the right panel, the $y$-axis is furthermore multiplied 
  by an $\epsilon$-dependent power law of radius chosen so that the
  inner profile appears flat and equal to unity (see Eq.~\ref{eq:profile}).
  Solid lines show the $a=1$ profiles, whereas dashed lines show the
  profiles at time $a=0.5$.}
\label{fig:scaled_profiles} 
\end{figure*}

The left panel of Fig.~\ref{fig:half_time_profiles} suggests that in the inner
part of the halo the $\beta$ parameter may approach an asymptotic value which
depends on the value of $\epsilon$.  \cite{2006NewA...11..333H} have proposed
that the equation $\beta(r)=-0.2({\rm d}({\rm ln}\rho)/{\rm d}({\rm ln} r + 0.8))$ relates
velocity anisotropy and density profile slope in the inner parts of dark
matter haloes (see also {\it e.g.}~\cite{2010MNRAS.402...21N}).  Using
Eq.~\ref{eq:profile} which is indeed valid in the inner part of our haloes, the
above relationship reduces to
\begin{equation}
\beta=0.2\left({9\epsilon\over (1+3\epsilon)}-0.8\right), \quad \epsilon > 0 \\ \\,
\label{eq:alpha-beta}
\end{equation}
which gives $\beta=\{0.17,0.2,0.23,0.25,0.29\}$ for
$\epsilon=\{0.4,0.5,0.6,0.7,0.9\}$.  These values are in good agreement with
those inferred from the left panel of Fig.~\ref{fig:half_time_profiles}, even
though our haloes form in a quite different (and ``nonuniversal'') manner from
CDM haloes.

\section{ The inner and outer scales: first pericentre and outer caustic radii}
\label{section:2scales}

In the previous section, we found approximate similarity behaviour for
nonspherical haloes formed by infall from power-law and almost spherically
symmetric initial conditions. As in the fully spherically symmetric similarity
solution of \cite{1985ApJS...58...39B}, the resulting nonlinear density
profile is a power law only at sufficiently small radii, but in our case the
transition between the infall and quasi-equilibrium regimes is more complex
than when the orbits are purely radial. As we have seen, there appear to be
two distinct scales: an inner scale at around $0.01 r_{\rm ta}$ and an outer
scale at around $0.1 r_{\rm ta}$.  In the last section we asserted that the
former can be associated with the first pericentric passage of infalling
particles, and the latter with the following apocentric passage, which occurs
at the outer caustic. Here we support these assertions by a more careful
study of the radial position of these orbital turning-points.

In the spherical similarity solution, the outer caustic is the border between
the one- and three-stream regions. Outside this radius all particles are
falling in for the first time, while within it many of them have passed
through the centre at least once. In Fig.~\ref{fig:caustic-ta} we show how the
radius of this outer caustic, evaluated directly from the similarity solution,
increases with $\epsilon$. Comparison with the
right panel of Fig.~\ref{fig:half_time_profiles} shows that the sharp change
in profile shape which delineates the inner boundary of the infall regime
in our simulations is indeed very close to this radius and depends on
$\epsilon$ exactly as predicted.
 
In Fig.~\ref{fig:half_time_profiles}, all radii are expressed in units of the
turnaround radius. Although this is the natural scale for models with
spherically symmetric initial perturbations of the kind studied here, it turns
out that scaling to the outer caustic radius (i.e. the position of the first
apocentre after turnaround) results in greater uniformity of the inner
nonlinear structure of our haloes as a function of $\epsilon$. We show this in
Fig.~\ref{fig:scaled_profiles} which may be compared with
Fig.~\ref{fig:half_time_profiles}, Note that in the right panel of
Fig.~\ref{fig:scaled_profiles}, the outer transition occurs slightly beyond
the nominal outer caustic radius. This is because we have rescaled using the
value for the spherically symmetric similarity solution, as plotted in
Fig.~\ref{fig:caustic-ta}. This is slightly smaller than caustic radius at the
end of our numerical bars, as is clearly visible in
Fig.~\ref{fig:density_map}.

Next, we consider the inner scale at around $0.01 r_{\rm ta}$ (see
Fig.~\ref{fig:half_time_profiles}).  We follow the trajactories of all
simulation particles and record the radial position of their first pericentric
passage. For all particles which pass first pericentre between $a=0.5$ and
$a=1$, we follow their radial position $r$ along their orbit and record its
first minimum, as well as the time when this minimum occurs. The latter allows
us to calculate the position of the outer caustic (according to the spherical
similarity solution) at the time of pericentric passage, and thus to measure
the ratio of these two lengths. A histogram of $r_{\rm peri}/r_{\rm outer\,\,
  caustic}$ is presented in Fig.~\ref{fig:peri_distance}. There is
considerable scatter in this ratio because pericentric radius depends strongly
on the angle between the infall direction of a particle and the long axis of
the bar. It is very small for infall along one of the principal axes, and it
maximises at angles well away from any of these axes. Nevertheless, there is a
well-defined ``typical'' pericentric distance which is $\sim 0.1 r_{\rm
  outer\,\, caustic}$ for all values of $\epsilon$.  This agrees well with the
inner transition scale of the $\beta$ and density profiles in
Figs.~\ref{fig:half_time_profiles} and~\ref{fig:scaled_profiles}.
  
\begin{figure*}
\center
{
\includegraphics[width=0.505\textwidth]{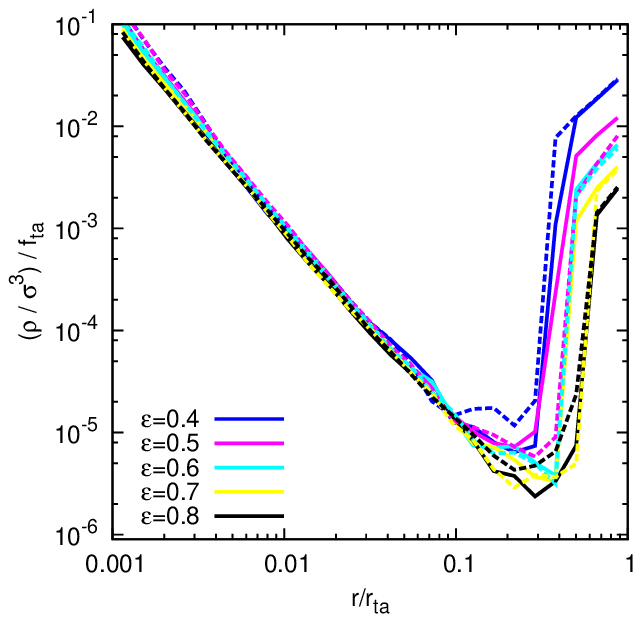}
\includegraphics[width=0.48\textwidth]{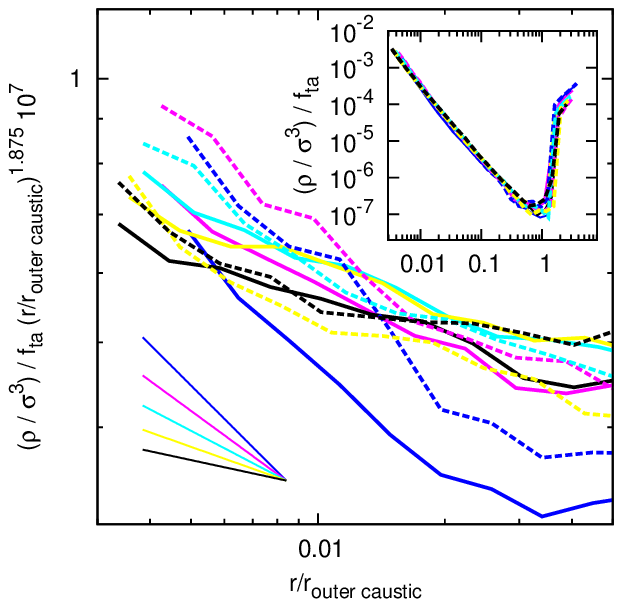}
}
\caption
{Left panel: Pseudo-phase-space density profiles for various $\epsilon$
 evaluated at $a=0.5$ (dashed) and $a=1$ (solid). The radial coordinate is
 scaled with $r_{\rm ta}$ and the pseudo-phase-space density with 
 $f_{\rm ta} \equiv \rho_bt^3/r_{\rm ta}^3$. With these scalings, the profiles do not
 change significantly with time, and, remarkably, are very similar and are
 close to a power law within the outer caustic for all $\epsilon$, even
 though though they differ in the infall region. There is no obvious feature
 at the radius of first pericentre. Right panel: Pseudo-phase-space density
 profiles of the inner parts of our haloes after rescaling the horizontal and
 vertical axes by $r_{\rm ta}/r_{\rm outer\,\, caustic}$ and by 
 $(r_{\rm  outer\,\, caustic}/r_{\rm ta})^3$ respectively. This scaling brings the
 overall profiles into quite close agreement within the outer caustic, as can
 be seen in the inset.  The slope of the inner profile nevertheless depends
 weakly on $\epsilon$. In the main panel we have additionally scaled by
 $r^{1.875}$, the inverse of the behaviour expected theoretically for
 $\epsilon=1$. Dashed lines show the pseudo-phase-space density profiles at $a=0.5$.
 The thin straight lines in the lower left indicate the slopes of the  
 analytic predictions of Eq.~\ref{eq:pseudo}.
 }
\label{fig:pseudo_profiles} 
\end{figure*}

\section{Pseudo-phase-space density profiles}
\label{section:psd}

A property of dark matter haloes with rather intriguing characteristics is the
pseudo-phase-space density.  This is defined as, $\rho/\sigma^3$, where
$\rho(r)$ and $\sigma(r)$ are the mass density and the (three-dimensional)
velocity dispersion averaged over a shell of radius $r$.  Even though neither
$\rho(r)$ nor $\sigma(r)$ is itself close to a power law, this particular
combination is very close to a power law for haloes formed from $\Lambda$CDM
initial conditions, and, remarkably, its power-law index is very similar to
that found in the spherically symmetric similarity solution of
\cite{1985ApJS...58...39B}. This curiosity was first pointed out by
\cite{2001ApJ...563..483T}, and since that time it has been investigated by
many authors who have found the behaviour to be quite robust, to extend over
more than three decades in radius, and to hold also for other kinds of initial
conditions \citep[see, for example,][]{2010MNRAS.tmp..718L,2010MNRAS.402...21N}).

In the framework of a true similarity solution with nonradial motions the
inner density profile must obey Eq.~\ref{eq:profile} for all $\epsilon>0$ and
$\beta(r)$ must become constant at small radii. Since the circular velocity
and the velocity dispersions must have the same scaling behaviour, the
pseudo-phase-space density is also a power law which is easily verified to be
\begin{equation}
{\rho\over \sigma^3} \propto  r^{-3(2+3\epsilon)/(2(1+3\epsilon))} \qquad {\rm for\,\,\, all\,\,\,}\epsilon \\ \\
\label{eq:pseudo}
\end{equation}
For the original similarity solutions of \cite{1984ApJ...281....1F} which
had purely radial orbits, this behaviour holds only for $\epsilon\geq 2/3$.
The specific case $\epsilon=1$ studied by \cite{1985ApJS...58...39B}
gives $\rho/\sigma^3\propto r^{-15/8}$.

We have directly evaluated the pseudo-phase-space density profiles of our
simulated haloes using logarithmic bins in radius and subtracting the mean
radial motion before evaluating the velocity dispersion. We show the results
in Fig.~\ref{fig:pseudo_profiles}.  In the left panel the radial coordinate is
normalised by the turnaround radius and the pseudo-phase-space density by the
characteristic value $f_{\rm ta} \equiv \rho_bt^3/r_{\rm ta}^3$. The very good
agreement between the curves for $a=0.5$ and for $a=1$ demonstrates that the
pseudo-phase-space density evolves self-similarly in time. This is no
surprise, given that we have already seen good scaling for the density and
velocity anisotropy profiles. More surprising is the fact that inside the
outer caustic radius the profiles show very little dependence on $\epsilon$,
either in slope or in amplitude. Furthermore, they are all good
approximations to a power law, and there is no obvious feature near the radius
of first pericentre despite the very strong features seen at this radius in
Fig.~\ref{fig:half_time_profiles}.  Once again scaling to the radius of the
outer caustic gives an even better overlap since it matches the break at
large radii, as can be seen in the inset in the right panel.

We focus on the innermost part of our haloes (the regions inside the first
pericentre) in the main plot of the right panel of
Fig.~\ref{fig:pseudo_profiles}. We again use the outer caustic radius to scale
the radii and pseudo-phase-space densities, and we additionally multiply the
pseudo-phase-space density by $r^{15/8}$ to take out the dominant trend.  The
power-law slopes expected for exact similarity scaling are indicated by thin
straight lines and are seen to be a good but not perfect fit to the simulated
behaviour. The differences are small enough that they can probably be ascribed
to residual time-dependent behaviour, an interpretation which is supported by
the differences between the $a=0.5$ and $a=1$ curves for each $\epsilon$.

Thus, unlike the velocity anisotropy and the density profiles, the
pseudo-phase-space density profile is close to a power law over the full
nonlinear extent of our haloes and it depends only very weakly on
$\epsilon$. Thus, to a good approximation it can be considered ``universal''.

\section{Conclusions}

We have shown that although the classic spherically symmetric similarity
solutions of \cite{1984ApJ...281....1F} and \cite{1985ApJS...58...39B} are
violently unstable to the radial orbit instability, evolution from the initial
conditions they presuppose gives rise to ellipsoidal, bar-like haloes which
nevertheless grow in an approximately self-similar way.

The nonlinear structure of these objects shows two characteristic radii.  The
outer caustic separates the infall and multistream regions and occurs at
approximately the same position as in the original spherically symmetric
similarity solutions. The second characeristic radius is about an order of
magnitude smaller and occurs at the typical first pericentre distance of the
infalling particle stream. It can be considered the outer edge of the
quasi-equilibrium region.  Relative to the turnaround radius, both radii
increase with the value of the similarity parameter $\epsilon$.

Both the density and velocity anisotropy profiles show strong features at the
first pericentre radius. Within this radius the density profile is
approximately a power law, $\rho\propto r^{-\gamma}$ with
$\gamma = 9\epsilon/(1+3\epsilon)$, and the velocity anisotropy is
approximately constant, $\beta = 0.2(\gamma - 0.8)$. At larger radii the
density profile becomes substantially steeper and the velocity anisotropy
rises steeply.

Despite the strong features in the density and velocity anisotropy
profiles at the first pericentre radius, the pseudo-phase-space density
profiles of all our haloes are close to power laws all the way out to
the outer caustic radius. Furthermore they depend only very weakly
on $\epsilon$ and are similar to the profiles of haloes formed from
$\Lambda$CDM initial conditions. Thus this profile seems remarkably
``universal''.

Our simulations were carried out at the Computing Centre of the
Max-Planck-Society in Garching. RM thanks French ANR OTARIE for support. We
thank Niayesh Afshordi, Ed Bertschinger, James Binney, Jacques Colin, Aaron
Ludlow and Julio Navarro for useful discussions.

\label{lastpage}
\end{document}